\documentclass[aps,prl,twocolumn,groupedaddress,floatfix]{revtex4}

\usepackage{graphicx}

\begin{document}


\title{A generalized inner and outer product of arbitrary multi-dimensional 
arrays using A Mathematics of Arrays (MoA)}


\author{James E. Raynolds} 
\address{College of Nanoscale Science and Engineering, University
at Albany, State University of New York, Albany, NY}
\author{Lenore M. Mullin}
\address{National Science Foundation, Arlington, VA}
\affiliation{}


\date{\today}

\begin{abstract}
An algorithm has been devised to compute the inner and outer product 
between two arbitrary multi-dimensional arrays A and B in a single 
piece of code.  It was derived using A Mathematics of Arrays (MoA) and the 
$\psi$-calculus.  Extensive tests of the new algorithm are presented for
running in sequential as well as OpenMP multiple processor modes.

\end{abstract}


\maketitle



%

\section{Introduction}

In this work we consider the efficient computation of inner and outer
products of arbitrary multi-dimensional arrays (tensors).  Our algorithm was 
presented in a previous work and was derived and expressed using the 
formalism known as A Mathematics of Arrays (MoA)~\cite{mul88}.  The routine 
maximizes data locality and computes both operations (inner and outer 
product) in a single piece of code.  In this work we emphasize computational
experiments and refer the reader to Ref~\citep{mul88} for details of the
formalism and the derivation.  

We now give a brief schematic discussion of the algorithm.  Using traditional
notation (as opposed to MoA), an outer product of two multi-dimensional
arrays (tensors) A and B, is given in terms of components of the result:
\begin{equation}
C_{ijkpqlm} = A_{ijkp} * B_{qlm}.
\label{outer}
\end{equation}
The MoA outer product is more general than given above in that the binary
operator $*$ (times) is generalized to be any binary operation (e.g. $+$,
$-$, $*$, $/$, etc.).

The MoA inner product is equivalent to a tensor contraction.  Working with 
the above arrays, we would write: 
\begin{equation}
D_{ijklm} = \sum_p A_{ijkp} * B_{plm},
\label{inner}
\end{equation}
where, as in the case of the MoA outer product, the binary operation $*$
(times) can be any binary operation.  From Eq.~\ref{inner} we can 
conclude two things: (1) the standard matrix multiply between two 
matrices $A$ and $B$ is a special case of the MoA inner product, and 
(2) the MoA inner product is intimately related to the MoA outer product
of Eq.~\ref{outer}.  It is therefore natural that both operations should 
be embodied in the same piece of code.

\begin{figure}
\includegraphics[height=7cm]{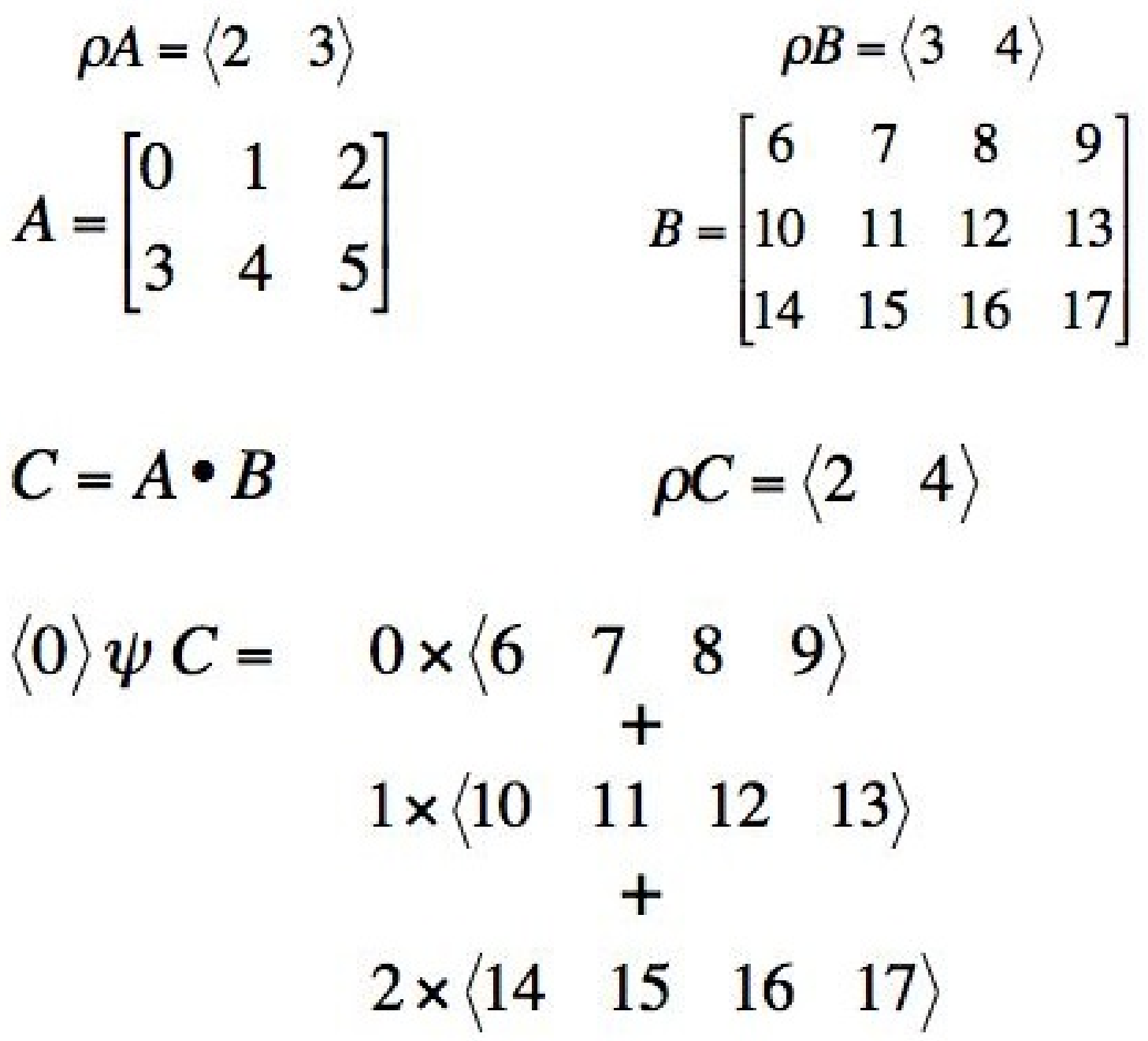}%
\caption{
\label{inner_fig}
Illustration of the general algorithm for the matrix multiply
(inner product) operation.  Data locality is maximized in that
an entire row of the result, $C$,  is computed at once while the elements
of the left array $A$ and right array $B$ are stored and accessed 
contiguously.  In this figure we illustrate how the first row (indicated
by $<0> \psi C$ is computed).
}
\end{figure}

Any arbitrary pair of arrays can be handled because of the generality of
the formalism and implementation.  The concept of array {\bf \em shape}
plays a key role.  The shape of an array is given by a vector whose
components give the lengths of the corresponding dimensions.  Thus an 
array input to the routine is described by the shape vector and a 
vector containing the elements of the array.  These concepts are illustrated
in Fig.~\ref{inner_fig}.  In this example, we take the arrays $A$ and $B$ to be
two dimensional (i.e. matrices).  The array $A$ has shape 
$\rho A = <\!2\;3\!>$ and $B$ has shape $\rho B = <\!3\;4\!>$.  In traditional
language we would say that $A$ is a $2\times 3$ matrix and $B$ is 
a $3\times 4$ matrix.

Data locality is maximized in that each row of the result (indicated by
$<0> \psi C$) is computed at a time and the elements of $A$ and $B$
are accessed contiguously.  In this figure we have illustrated some of the
notational devices of MoA.  In this formalism, the $\psi$ operator selects
components and subarrays of a given array with the use of an index vector.
In this case we select the {\em zero'th} row of $C$ with the operation
$<0> \psi C$.  We see that each row of $B$ is multiplied by an element of 
$A$ and then added to the next row of $B$ multiplied by an element $A$.
This formulation of the inner product might seem simple but is actually 
quite subtle in the general case of two arbitrary multi-dimensional arrays.
This way of organizing the operations leads to significant performance
gains as will be demonstrated in the numerical tests to be described below.

In the rest of this paper we present performance tests of our routine for 
the computation of the standard matrix multiply in comparison with a benchmark 
routine (dgemm.f) taken from the BLAS library.  We present tests of both
sequential and OpenMP parallel implementations.  We find that our 
routine is either competative or outperforms the BLAS routine. 

As discussed more extensively below, our goal is to demonstrate the 
advantages of our design methodology using MoA.  We don't claim to have 
established the the best matrix multiply and in no way do we wish to 
enter such a competition.  Indeed the matrix multiply has been extensively
studied~\cite{dongarra88a,dongarra88b,dongarra90a,dongarra90b,blackford02,
whaley01,gunnels01,gunnels05,gunnels06,vandegeijn97,
low05,bientinesi05a,bientinesi05b,lawson79a,lawson79b,anderson99} and we 
defer to the experts for those searching for the best matrix multiply.  For 
our OpenMP version we also make no attempt at optimization.  We simply adopt a 
``poor man's" parallelism by wrapping sequential code with the simplest
OpenMP statements (not even specifying a ``chunk" size, for example).  The
point is that we take an ``off the shelf" sequential benchmark, the 
standard BLAS routine dgemm.f and compare it with our generalized inner
and outer product code for the limited case of matrix multiply and we find
competative results without any optimizations other than the fortran 
compiler options $-O0$, $-O1$, $-O2$ and $-O3$.  For high performance 
parallel matrix multiply we again defer to the experts cited above as well
as those in Refs.~\citep{bentz05,addison,santos03,bilmes97,demmel05,bernsten89,
choi94,lederman93a,li01,irony04}.

\section{Numerical experiments}

\subsection{Computational Environment}
A series of sequential and multi-processor tests were carried out for 
the MoA routine in comparison with the standard BLAS dgemm.f.  
The key code fragments are presented in the Appendix.
A dedicated, non-shared, computational environment was used on the 
5,120 processor machine ``jaws" at the Maui High-Performance Computing Center.  The following
information is quoted from the website (www.mhpcc.hpc.mil):

``Jaws is a Dell PowerEdge 1955 blade server cluster comprised of 5,120 
processors in 1,280 nodes. Each node contains 2 Dual Core 3.0 GHz 64-bit 
Woodcrest CPUs, 8GB of RAM, and 72GB of local SAS disk.  Additionally, 
there is 200TB of shared disk available through the Lustre filesystem. The 
nodes are connected via Cisco Infiniband, running at 10Gbits/sec (peak).
Jaws has a peak performance of 62400 GFlops, and LINPACK performance of 
42390 GFlops."

In the following we will present results for our new routine run in 
sequential and OpenMP multi-threaded tests.

\subsection{Tests for matrix-matrix multiply}

Our object of study is the computation of the matrix multiply
$C = A*B$ where we consider $C$ to be a $m \times n$ matrix, where 
$m$ is the number of processors (threads) and $n$ is an integer power of 
$2$ and is varied from the smallest to largest sizes that can be accomodated.
The matrix $A$ has dimensions $m \times \ell$ while $B$ has dimensions
$\ell \times n$.  For these tests we keep $\ell$ fixed at the value
$\ell = 128$.

There are two performance metrics of interest in this study: (1) the 
``time per thread" and (2) the ``total time".  For a multi-thredded job
the ``time per thread" is simply the total time for the job to run.  Note
that a job with $m$ threads is dealing with a problem size that is $m$
times as large as the problem considered on $1$ thread.  Thus if there were
no communcation costs we would expect the curve of ``time per thread" vs.
$n$ to be the same, independent of the number of threads $m$.

In some cases we wish to consider a fixed problem size and see how long it
takes on $m = 1$, $2$, $3$, and $4$ processors.  In this case we take the 
curves discussed in the previous paragraph and scale the $x$ axis 
(i.e. problem ``size") of each curve by multiplying by the corresponding 
number of processors $m$.  This type of plot should explicitly show the 
benefit of parallelism (if there is one) if the curve for $m$ threads
is below that for $1$ thread.

\subsection{Sequential tests}

In a first series of numerical experiments we tested the MoA routine and 
the BLAS dgemm.f routine in sequential mode in a dedicated non-shared 
batch environment.  Perl scripts were used in each case to compile the
routine for a given value of $n$ and then the job is timed.  This process
is repeated three times for each $n$ and the timings were averaged.  As 
reproducibility is a key concern, we also repeated several tests on 
different days of the week to make sure there were no substantial 
fluctuations.  In all cases
tested we found essentially identical results.  From these careful 
considerations we conclude that all results presented in this work are
{\bf \em reproducible}.

Our initial interest was in determining the effect of
compiler options on the performance of our routine and the BLAS routine.  We 
thus ran our experiments with the four optimization flags: 
$-O0$ (no optimization), $-O1$, $-O2$ and $-O3$ in four separate tests 
respectively.  We used the Intel Fortran compiler ``ifort" that was supplied
with the machine.  In all cases we found the compiler option $-O1$ to give 
the best performance.  Thus in all results to be presented, we assume the
compiler option $-O1$ to be in effect.  Interestingly, for the MoA routine,
we find a benefit on going from $-O0$ to $-O1$ but then no difference between
$-O1$, $-O2$, and $-O3$. In constrast, however, for the BLAS routine we find
the speed to increase upon going from $-O0$ to $-O2$ and then to increase
again upon going from $-O2$ to $-O1$ while the results for $-O2$ and 
$-O3$ were essentially identical.

Comparison of the MoA routine vs. the BLAS routine dgemm.f run in sequential
mode are presented in Fig.~\ref{fig1}.  We see that the results are 
comparable with a slight benefit given by the MoA routine for small sizes.
For the largest sizes that fit in real memory, the BLAS routine outperformes
the MoA routine but for sizes requiring virtual memory the results are
essentially the same.  We will see that the superior performance of the 
BLAS routine at largest sizes is lost when we go to multiple threads 
using OpenMP.

\subsection{Multiple-thread OpenMP}

Our next set of computational experiments were performed using OpenMP
multiple threads.  On this machine (see description in first section)
each node contains four processors (two dual cores) and so we 
restrict our attention in 
this series of experiments to $m = 1$, $2$, $3$, and $4$ threads.  Making
the transition from a sequential piece of code to open OpenMP is achieved
by wraping the seqential algorithm, in each case with simple OpenMP directives.
No attempt was made to optimize the parallel performance of either routine.

Our goal in the multi-threaded tests to be discussed herein is as follows.
We are proceeding from a general, mathematically-based design methodology.  
Thus although our code was not designed to specifically exploit the 
multi-threading capabilities of this machine, we achieve impressive results
in comparison with the BLAS benchmark.  Again, we emphasize the design
methodology. Our approach is completely mechanizable from the ONF (Operational
Normal Form). That is given start, stop, stride, count, we can
instantiate the software at any 
level of memory~\cite{rmhr,mullin-raynolds-book,Mul03,mul91,cpc}.  We are not trying to claim that we have achieved the fastest
multi-threaded matrix multiply.  Nor are we comparing our results against
a BLAS routine that has been designed for multi-processor, multi-threaded
hardware.  That is not our goal, but rather to argue the merits of a
design methodology that consistently leads to efficient implementations 
by eliminating temporaries and exploiting data locality.

\begin{figure} 
\includegraphics[height=7cm]{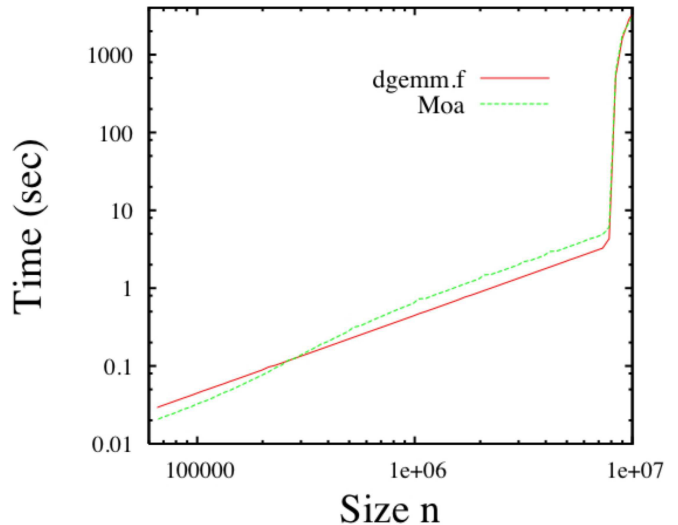}%
\caption{
\label{fig1} 
Best sequential MoA result compared with the best
sequential BLAS result (flag -O1).}
\end{figure}

\begin{figure} 
\includegraphics[height=6cm]{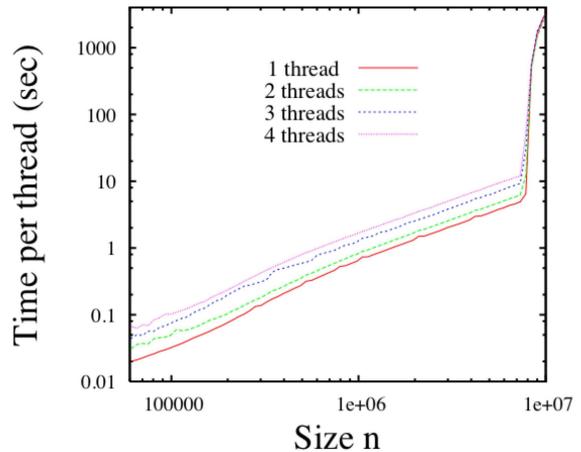}%
\caption{
\label{fig2} 
Comparison of the {\bf \em time/thread} for 
threads $m = 1$, $2$, $3$ and $4$ assuming the fastest compiler option in 
each case (i.e. -O1) for the MoA routine.  
Note, the problem size is proportional to the number of threads $m$.  Thus 
the differences between the four curves represent communication costs.}
\end{figure}

Figure~\ref{fig2} presents results for $m = 1$, $2$, $3$, and $4$ OpenMP
threads for the MoA routine.  We plot the {\bf \em time/thread} for 
each job.  In other words this is the total time for the job to
run with the size of the problem proportional to the number of threads $m$.
This metric illustrates the communication cost associated with multiple 
threads because, in the absence of communication cost (i.e. in a situation
of ``perfect parallism") the time/thread vs. $n$ (i.e. the number of 
columns of $C = AB$) would be independent of the number of threads $m$.

\begin{figure} 
\includegraphics[height=6cm]{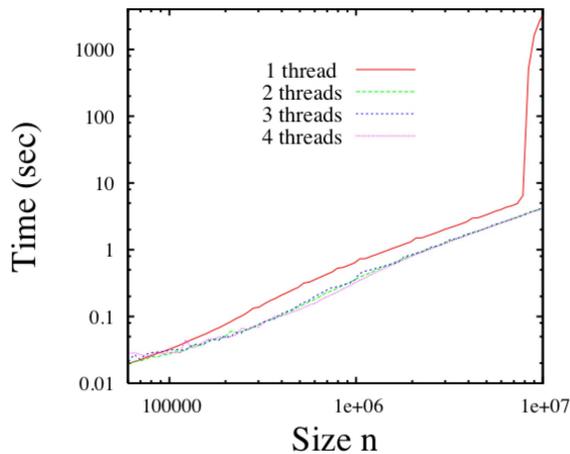}%
\caption{
\label{fig3} 
Comparison of the {\bf \em total time} for threads 
$m = 1$, $2$, $3$ and $4$ assuming the fastest compiler option in each 
case (i.e. -O1) for the MoA routine.   These curves were obtained from the 
ones of Fig.~\ref{fig2} by rescaling the $x$-axis of each curve by 
multiplying by the corresponding number of processors $m$.  Thus, in this 
case the $x$ axis ($n$) represents the total problem size.  Note for this 
range of $n$, there is benefit in going from $m = 1$, to $2$ but there is no 
net benefit in using $m = 3$ and $4$.
}
\end{figure}

In Fig.~\ref{fig3} we emphasize the net benefit of the use of multiple threads
by considering the ``total time" vs. the size of the problem.  In other 
words, in this case, for each value of $m$ (i.e. the number of threads) 
we scale the $x$-axis of the ``time/thread" plot illustrated in 
Fig.~\ref{fig2} by $m$.  
Thus for a given value of $n$, if the curve for a given number of threads
lies below that for $m = 1$, there is a net benefit to using multiple
threads.  We see that in this series of tests, there IS a net benefit to 
the use of $m = 2$ threads but there is no net benefit for $m > 2$ threads.

\begin{figure} 
\includegraphics[height=6cm]{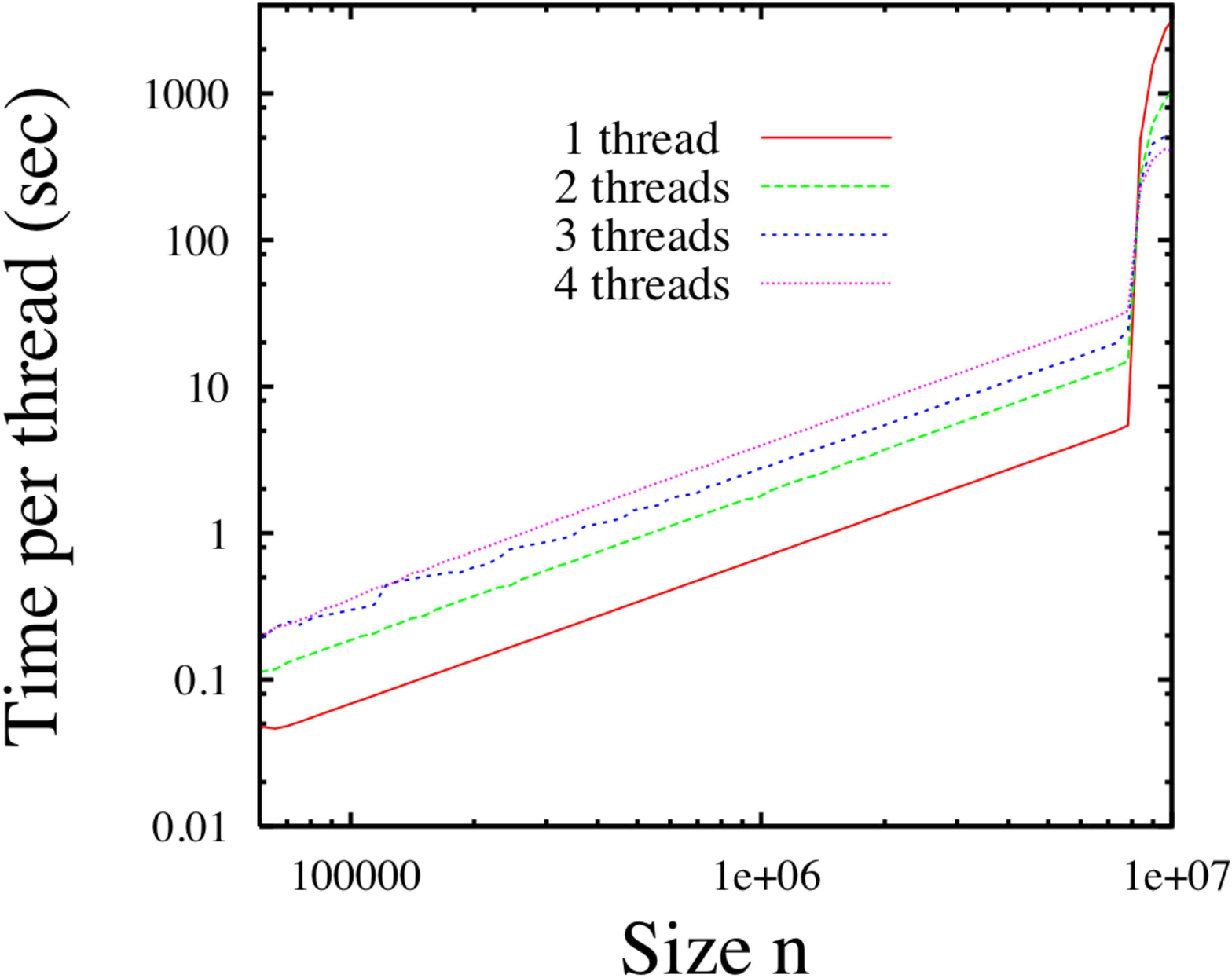}%
\caption{
\label{fig4}
Comparison of the {\bf \em time/thread} for threads $m = 1$, $2$,
$3$ and $4$ assuming the fastest compiler option in each case (i.e. -O1)
for the BLAS routine dgemm.f.  Note, the problem size is proportional to the 
number of threads $m$.  Thus the differences between the four curves represent 
communication costs.
}
\end{figure}

In Fig.~\ref{fig4} we plot the ``time/thread" for the BLAS routine with
$m = 1$, $2$, $3$ and $4$ threads.  The curves look similar to those for
the MoA routine of Fig.~\ref{fig3} but as we will see the following figure
there is a fundamental difference.

\begin{figure} 
\includegraphics[height=6cm]{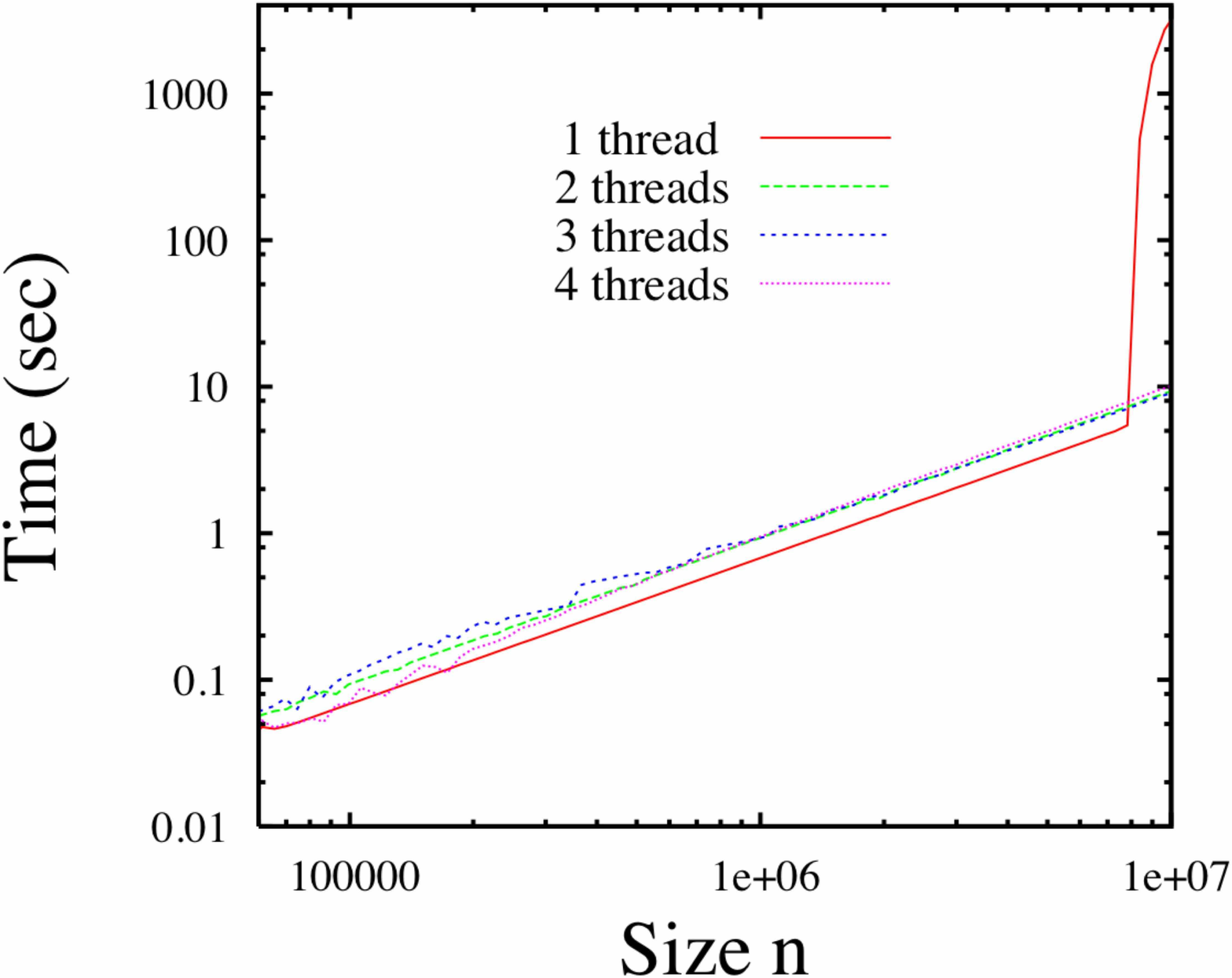}%
\caption{
\label{fig5}
Comparison of the {\bf \em total time} for threads
$m = 1$, $2$, $3$ and $4$ assuming the fastest compiler option in each
case (i.e. -O1) for the BLAS routine dgemm.f.   These curves were obtained 
from the ones of Fig.~\ref{fig4} by rescaling the $x$ axis of each curve 
by multiplying by the corresponding number of processors $m$.  Thus, in this 
case the $x$ axis ($n$) represents the total problem size.  Note for this 
range of $n$, there is benefit in going from $m = 1$, to $2$ but there is no
net benefit in using $m = 3$ and $4$.
}
\end{figure}

The results for the ``total time" vs. $n$ for the BLAS routine are presented
in Fig.~\ref{fig5}.  The results of Fig.~\ref{fig5}, for the BLAS routine
are fundamentally different from those for the MoA routine of Fig.~\ref{fig3}
in that, while all the $m > 1$ curves of Fig.~\ref{fig3} lie {\bf \em below}
the curve for $m = 1$, in Fig.~\ref{fig5} we see all $m > 1$ curves
lie {\bf \em above} the $m = 1$ result. {\bf Thus for this series of experiments
there is no net benefit to the use of multiple threads for the BLAS routine.}

\begin{figure} 
\includegraphics[height=6cm]{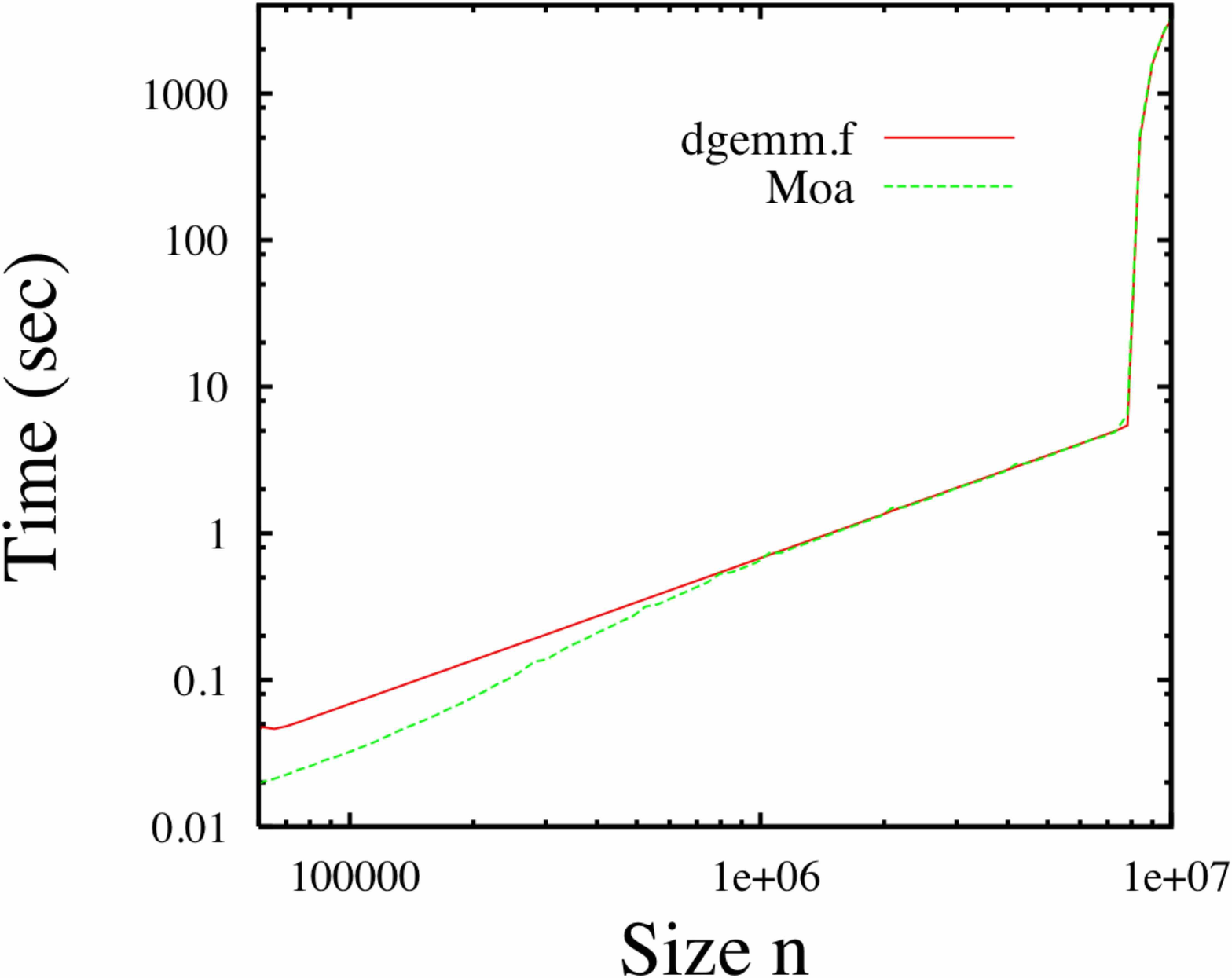}%
\caption{\label{fig6} Comparison of the MoA routine with the BLAS routine
dgemm.f for one thread. The MoA routine is superior for smaller sizes
while the two are equivalent for large sizes.
}
\end{figure}

In the next four figures we compare the ``time/thread" for the MoA routine,
directly with the BLAS routine.  In Fig.~\ref{fig6} we compare the one-thread
result for the MoA routine with the BLAS routine.  We find that the MoA
result out performs the BLAS routine for small matrix sizes and is equivalent
to that of the BLAS routine for the largest sizes.  Note that this figure
should be directly compared with Fig.~\ref{fig1} for the sequential 
runs.  We see that, although BLAS had the advantage for the largest sizes
when running in sequential mode, the advantage is lost when going to 
$m = 1$ thread using OpenMP.

\begin{figure}
\includegraphics[height=6cm]{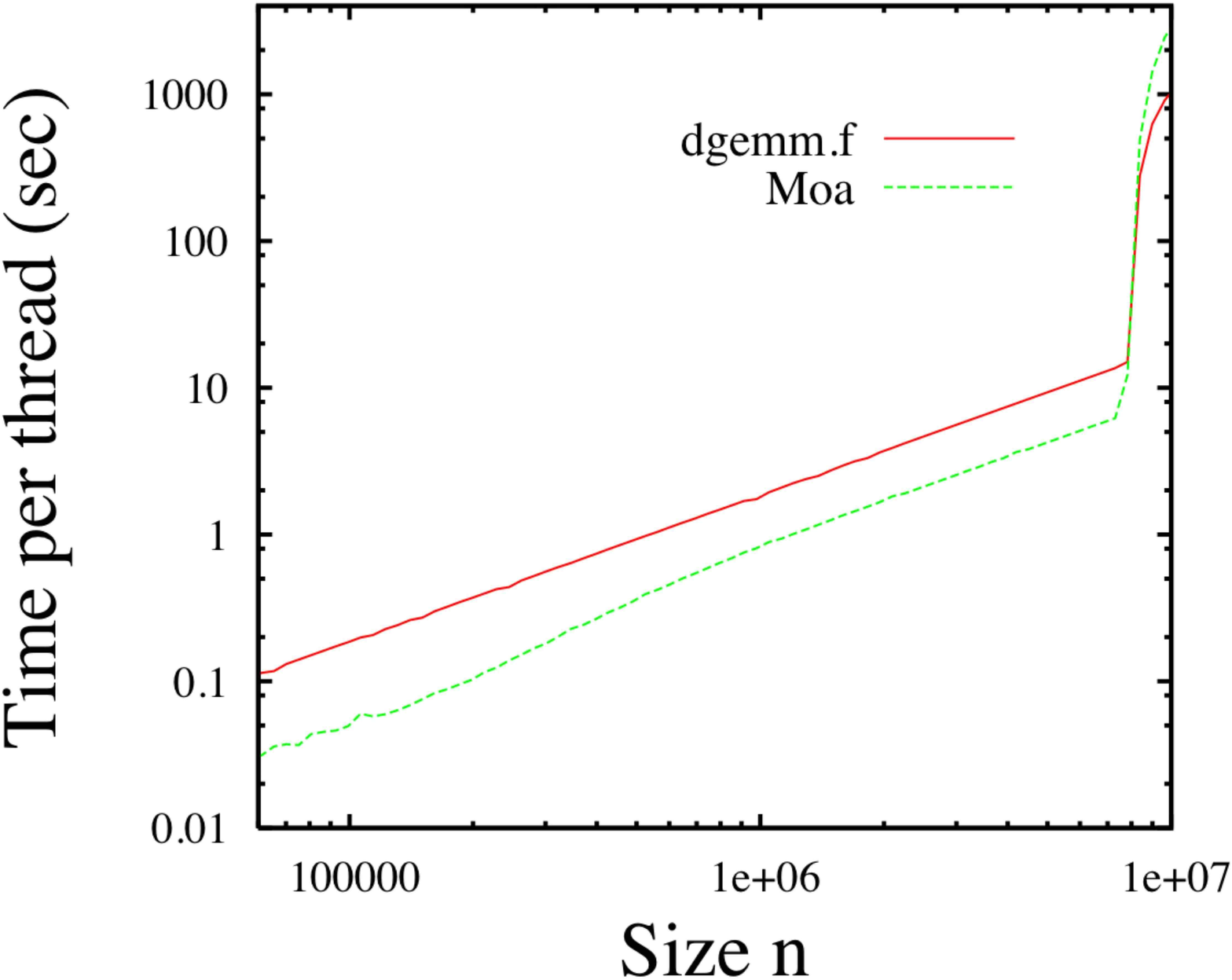}%
\caption{\label{fig7} Comparison of the MoA routine with the BLAS routine
dgemm.f for two threads. The MoA routine is superior for all sizes
that fit in real memory.
}
\end{figure}

For Figs.~\ref{fig7}, through~\ref{fig9} we see that the MoA routine 
consistently out performs the BLAS routine for all sizes that fit 
into main memory.  The BLAS routine out performs the MoA routine for sizes
that only fit in virtual memory.  The success of the MoA routine is due to
the data locality of the algorithm's contiguous 
access of all arguments: an output with two inputs.

\begin{figure} 
\includegraphics[height=6cm]{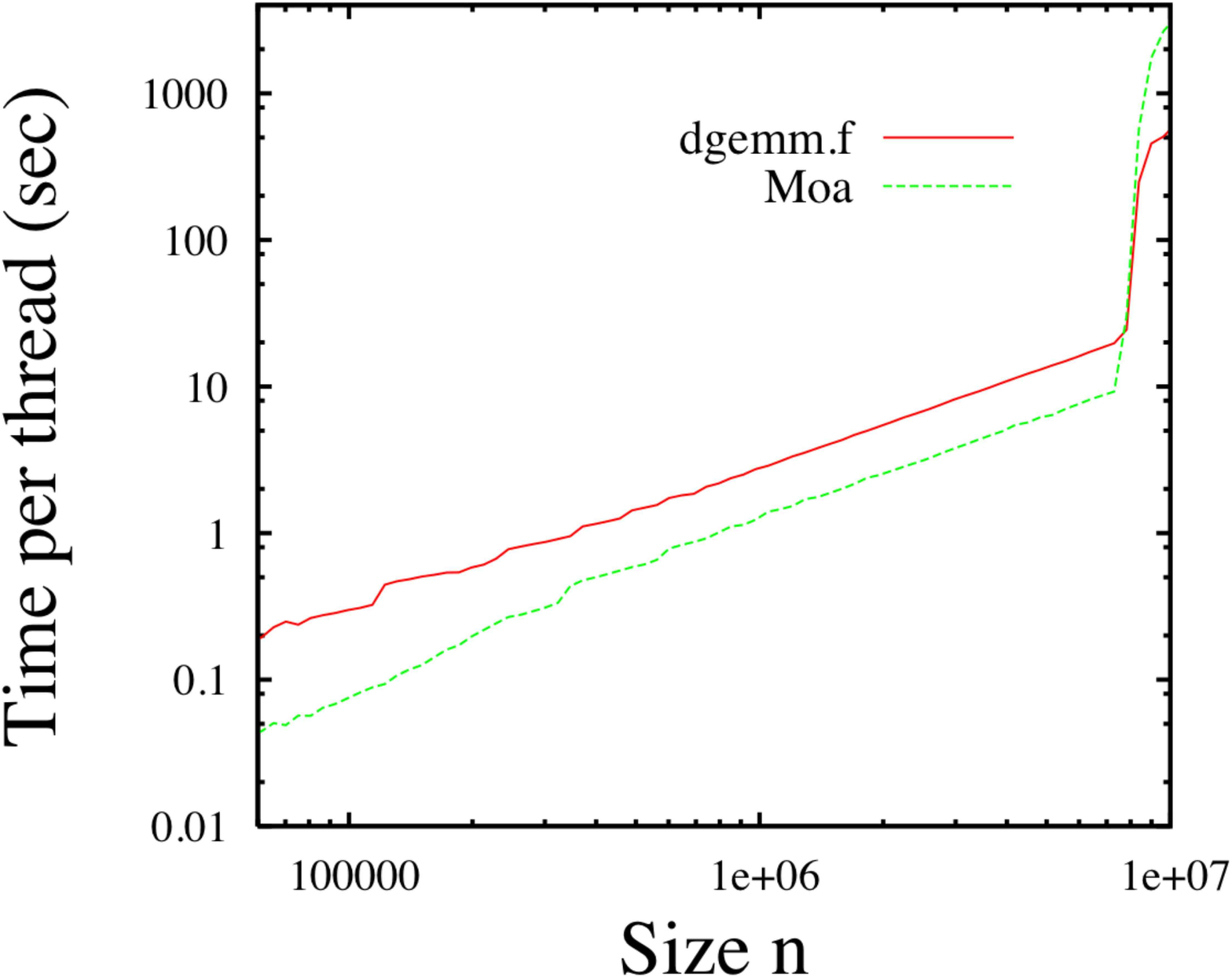}%
\caption{\label{fig8} Comparison of the MoA routine with the BLAS routine
dgemm.f for three threads. The MoA routine is superior for all sizes
that fit in real memory.
}
\end{figure}

\begin{figure} 
\includegraphics[height=6cm]{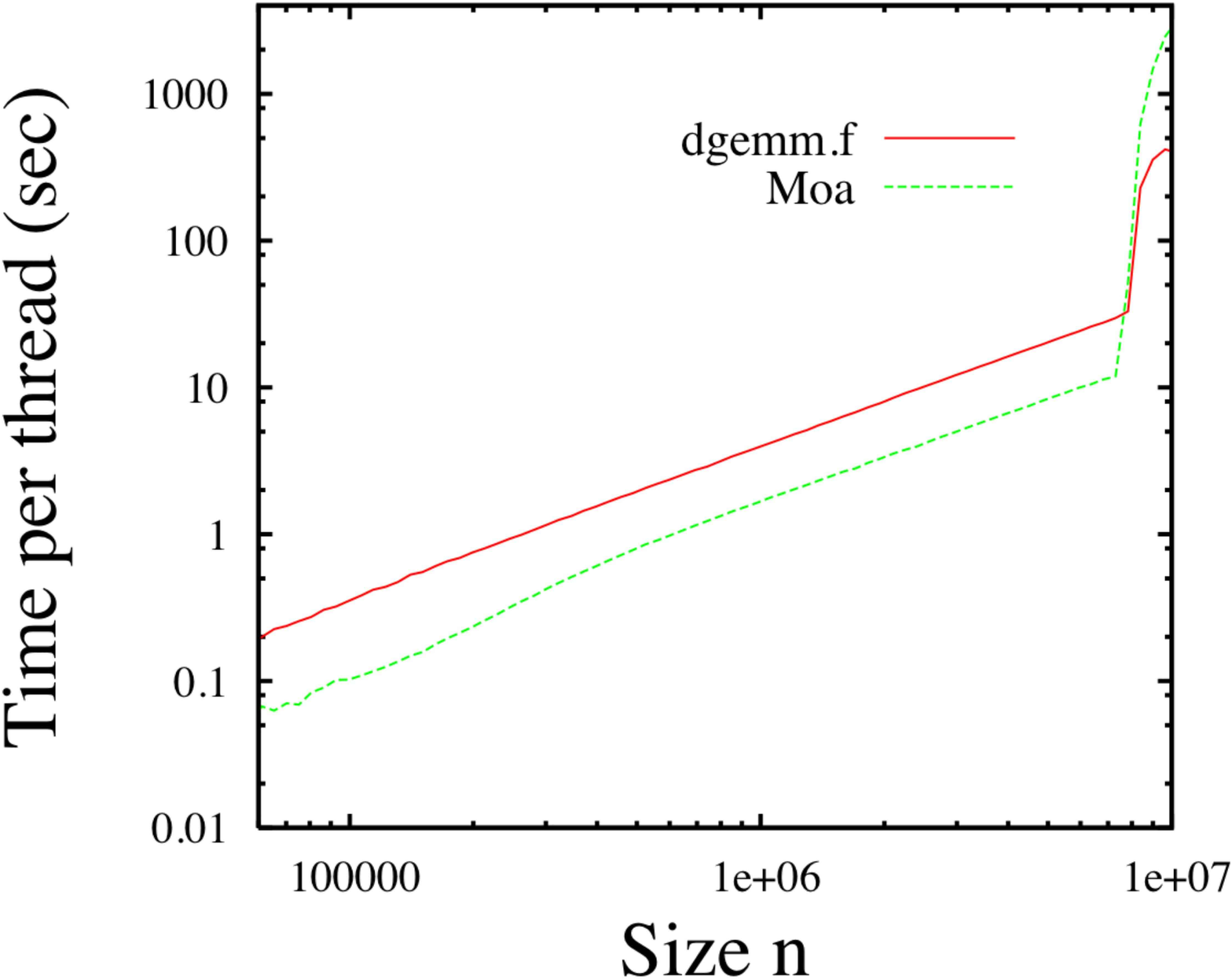}%
\caption{\label{fig9} Comparison of the MoA routine with the BLAS routine
dgemm.f for four threads. The MoA routine is superior for all sizes
that fit in real memory.
}
\end{figure}

\begin{figure} 
\includegraphics[height=6cm]{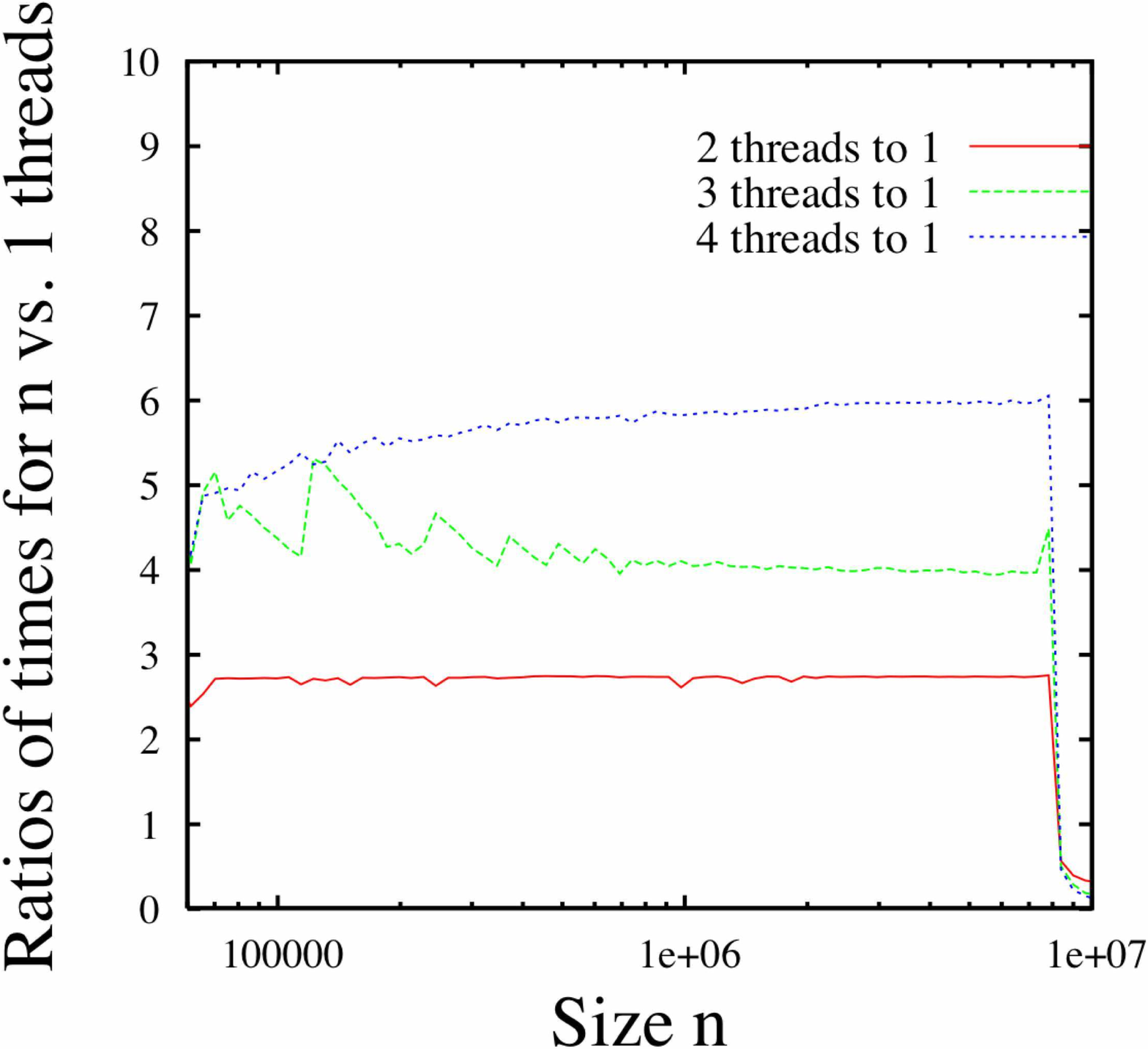}%
\caption{\label{fig10} Ratio of the time/thread, for a given number of 
threads, $m$, to that for one thread for the BLAS routine dgemm.f.  This 
metric allows one to judge the benefit of parallelism.  If there were no 
communication costs, such a ratio would be unity indicating ``perfect 
parallelism".  If this ratio is greater than $m$ (as in this figure) then
the overhead is more expensive than running $m$ jobs sequentially 
(``perfect sequentialism"(?)).  These results for this (unoptimized) routine
indicate no benefit to the use of multiple threads.
}
\end{figure}

\begin{figure} 
\includegraphics[height=6cm]{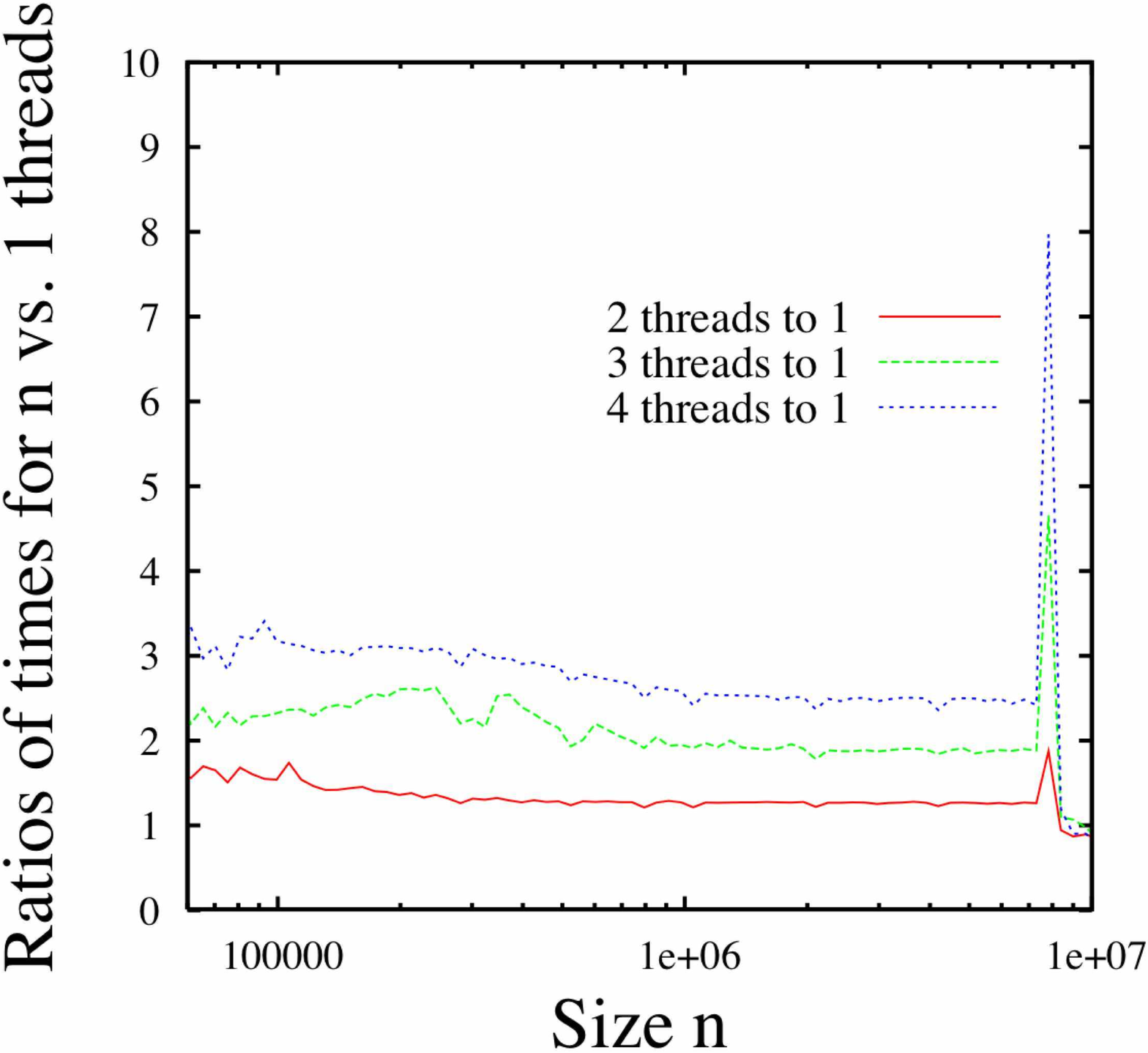}%
\caption{\label{fig11} Ratio of the time/thread, for a given number of threads, 
$m$, to that for one thread for the MoA routine.  This metric allows one to 
judge the benefit of parallelism.  If there were no communication costs, 
such a ratio would be unity indicating ``perfect parallelism".  If this ratio 
is greater than $m$ then the overhead is more expensive than running $m$ 
jobs sequentially (``perfect sequentialism"(?)).  For this (MoA) routine
the ratio is intermediate between unity and $m$, in each case, indicating a 
net benefit to the use multiple threads.
}
\end{figure}

Another way to characterize the data we have considered so far is as 
follows.  We consider the ``time/processor" for each number of threads
$m$ divided by the result for a single thread.  Such results are
presented in Figs.~\ref{fig10} and~\ref{fig11} for the BLAS routine
and the MoA routine respectively.  As argued previously, such a ratio
should illustrate the effects of communication costs.  If there were
no communication cost, each ratio would be unity.  Next, we consider
the notion that if the ratio is greater than $m$, there is no net benefit
to using multiple threads as this would indicate that the job was more
expensive than $m$ sequential jobs.  For these (unoptimized) tests we
conclude from Figs.~\ref{fig10} and~\ref{fig11} that there is {\bf \em
 no net benefit} to the use of OpenMP multiple threads for the BLAS
routine while there IS a net benefit to such use for the MoA routine.

Again, we emphasize that this result is not definitive for establishing
a matrix multiply that is superior to BLAS.  Indeed there are BLAS routines
(and others) that are optimized for multiple threads and multiple 
processors~\cite{addison,bentz05,santos03,bernsten89,chaterjee99,cherkassky88,
choi94,demmel93,fox87,lederman93a,li01,irony04,vandegeijn97}.  
We only
emphasize the quality of our results as an advertisement for our methodical
software design approach that exploits data locality as a fundamental 
principle.

\section{Conclusion}

We have presented numerical tests of a generalized inner and outer 
product routine applicable to arbitrary multi-dimensional tensors specified
at run time. Our algorithm computes either operation in a single piece of 
code.  In this work we have focused on the limited case of matrix-matrix
multiplication and have tested its performance for matrices from small sizes
to the largest that can be possibly accomodated.  As a benchmark reference
we compare our results with the standard BLAS dgemm.f routine.  We find 
that our routine is competative or out performs the BLAS routine.  We have
also presented tests of our routine using OpenMP parallelization without
any machine specific optimizations (other than the compiler options $-O0$,
$-O1$, $-O2$ and $-O3$).  Again we find competative performance.  As stated
earlier, our goal is not to claim the best matrix multiply routine but rather
to give definative tests of our routine for a well studied example: matrix
multiply.  Rather, we wish to emphasize the generality of our routine that
can compute the inner and outer product between two arbitrary multi-dimensional
arrays, as specified at run time, in a single piece of code.


%



\appendix
\section{Code fragments for the numerical experiments}

This appendix presents the key code fragments used in the testing of 
the BLAS routine (Fig.~\ref{blas_code}) and the MoA routine 
(Fig.~\ref{moa_code}) 

\begin{figure}
\begin{verbatim}
c$OMP do private(i,j,l)
              DO 90 J = 1,N
                  IF (BETA.EQ.ZERO) THEN
                      DO 50 I = 1,M
                          C(I,J) = ZERO
   50                 CONTINUE
                  ELSE IF (BETA.NE.ONE) THEN
c  do 60 vectorized
                      DO 60 I = 1,M
                          C(I,J) = BETA*C(I,J)
   60                 CONTINUE
                  END IF
                  DO 80 L = 1,K
                      IF (B(L,J).NE.ZERO) THEN
                          TEMP = ALPHA*B(L,J)
c  do 70 vectorized
                          DO 70 I = 1,M
                              C(I,J) = C(I,J) + TEMP*A(I,L)
   70                     CONTINUE
                      END IF
   80             CONTINUE
   90         CONTINUE
c$OMP end do nowait

\end{verbatim}
\caption{
\label{blas_code}
Key code fragment used in the benchmark tests of the BLAS routine.  The
first and last lines are the only OPenMP directives used in these tests
and are the same as those used in the MoA tests.
}
\end{figure}

\begin{figure}
\begin{verbatim}
c$OMP do private(k,i,l,j)
      do 100 k=0,(nthreads-1)
        do 120 i=k,(noproc-1),nthreads
          do 140 l=0,(rowsinred-1)
            do 160 j=0,(elsinop-1)
              RESADDR(1 + (i*restride) + j) =
     +              RESADDR(1 + (i*restride) + j) +
     +                 LADDR(1 + l + (i*lstride))*
     +                    RADDR(1 + (l*rstride) + j);
160         continue
140       continue
120     continue
100   continue
c$OMP end do nowait

\end{verbatim}
\caption{
\label{moa_code}
Key code fragment used in the tests of the MoA routine.  The
first and last lines are the only OPenMP directives used in these tests and
are the same as those used in the BLAS tests.
}
\end{figure}


\bibliography{paper}

\end{document}